\documentclass{aa}
\usepackage{graphicx,natbib}

\usepackage{txfonts}
\begin{document}
\title{Multiband Photometry of the Blazar PKS\,0537$-$441: \\ A Major Active State in December 2004 - March 2005}

\author{A. Dolcini\inst{1}, S. Covino\inst{2}, A. Treves\inst{1}, E. Palazzi\inst{3}, E. Pian\inst{4}, E. Molinari\inst{2}, G. Chincarini\inst{2,5}, F.M. Zerbi\inst{2}, M. Rodon\'o\inst{6,7}, V. Testa\inst{8}, G. Tosti\inst{9}, F. Vitali\inst{8}, L.A. Antonelli\inst{8}, P. Conconi\inst{2}, G. Cutispoto\inst{6}, A. Monfardini\inst{10,11},  M. Stefanon\inst{12}, P. D'Avanzo\inst{1,2}, J. Danziger\inst{5}, A. Fernandez-Soto\inst{13}, E. Meurs\inst{14}}

\offprints{S. Covino, covino@merate.mi.astro.it}

\institute{ 
Universit\`a dell'Insubria, Dipartimento di Fisica e Matematica, via Valleggio 11, 22100, Como, Italy.
\and 
INAF, Osservatorio Astronomico di Brera, via E. Bianchi 46, 23807 Merate (Lc), Italy.
\and 
IASF/CNR, Sezione di Bologna, Via Gobetti 101, 40129 Bologna, Italy.
\and 
INAF, Osservatorio Astronomico di Trieste, Via G. B. Tiepolo 11, 34131 Trieste, Italy.
\and 
Universit\`a degli studi di Milano-Bicocca, Dipartimento di Fisica, Piazza delle Scienze 3, 20126 Milan, Italy.
\and 
INAF, Osservatorio Astrofisico di Catania, Via S. Sofia 78, 95123 Catania, Italy.
\and 
Dipartimento di Fisica e Astronomia Universit\`a di Catania, v. S. Sofia 78, 95123, Catania, Italy.
\and 
INAF, Osservatorio Astronomico di Roma, Via Frascati 33, 00040 Monteporzio Catone, Italy.
\and 
Dipartimento di Fisica e Osservatorio Astronomico, Universit\`a di Perugia, Italy.
\and 
Astrophysics Research Institute, Liverpool JMU, United Kingdom.
\and 
ITC-IRST and INFN, Trento, Italy.
\and 
European Southern Observatory, Karl-Schwarzschild-Str. 2, 85748 Garching bei M\"unchen, Germany.
\and 
Observatori Astronomic, Universitat de Valencia, Aptdo. Correos 22085-Valencia, 46071, Spain.
\and 
Dunsink Observatory, Castleknock, Dublin 15, Ireland. 
}

\date{}

\abstract{
Multiband $VRIJHK$ photometry of the Blazar PKS\,0537$-$441 obtained with the REM telescope from December 2004 to March 2005 is presented. A major period of activity is found with more than four magnitudes variability in the $V$ filter in 50 days and of 2.5 in 10 days. In intensity and duration the activity is similar to that of 1972 reported by \citet{Egg73}, but it is much better documented. No clear evidence of variability on time-scale of minutes is found. The spectral energy distribution is roughly described by a power-law, with the weaker state being the softer.
\keywords{(Galaxies:) active -- (Galaxies:) BL Lacertae objects: PKS\,0537$-$441}
}

\authorrunning{Dolcini et al.}

\titlerunning{Multiband Photometry of PKS\,0537$-$441}

\maketitle
%

\section{Introduction}
\label{sec:intro}

Blazars are Active Galactic Nuclei (AGN) characterised by high variability in all bands and high polarisation.  Generally, the variability is wavelength dependent, with larger amplitudes at higher frequencies and a tendency to hardening with increasing intensity \citep[for variability of Blazars see e.g.][]{Ulr97}. Optical variability may be very complex, with episodes of month to  year duration, involving changes of several magnitudes. Intranight episodes of a fraction of a magnitude are not rare.

The photometry of Blazars has started some 30 years ago with the very discovery of the class. However, long and systematic campaigns, capable of detecting short time-scale variability, are relatively uncommon \citep[see][and references therein]{TS99}. Basic problems are, on the one hand, the availability of telescope observing time for this kind of programs, on the other hand the heavy human commitment required by the long observing runs and by the data analysis. Automatic (or robotic) telescopes represent a class of instruments which can overcome the difficulties mentioned above possibly together with global networks of observatories such as the WEBT (Whole Earth Blazar Telescope\footnote{http://www.to.astro.it/blazars/webt/}).

In this spirit a systematic program of photometry of a number of southern Blazars was started with the Rapid Eye Mount (REM) telescope, which was constructed by an international consortium \citep{Chinc03} and designed to also complement the Swift mission \citep{Geh04}.

Here we report results on the Blazar PKS\,0537$-$441, obtained from December 2004 to March 2005, at a stage when the software of the photometric facility providing in real-time the quick-look of the analysis was not yet fully operational. This paper is organised as follows: in Sect.\,\ref{sec:pks} a description of our target source is given; in Sect.\,\ref{sec:data} we briefly present the instrumentation and data analysis; results and discussions are reported in Sect.\,\ref{sec:disc}.

\section{The Target: PKS\,0537$-$441}
\label{sec:pks}

PKS\,0537$-$441 ($z=0.896$) is one of the most luminous and variable Blazar, at all frequencies. It has been observed repeatedly from the radio to the  gamma-rays. The overall spectrum in different intensity states has been discussed in detail by \cite{Pian02}. We refer to this paper for literature on the source. The infrared to gamma-ray spectrum is well represented by synchrotron self-Compton models with a contribution from inverse Compton scattering off emission line photons at the highest energies.

Optical photometry and polarimetry were carried out in recent years by \citet{Rom00,Rom02} and by \citet{Andr05}, who found the source at $V = 16.2 - 15.5$, with a polarisation of 10\% and in a modest variability state. An active state is reported by \citet{HW96} with a V variability of $\sim 0.3$\,mag in two nights.

\section{Telescope and Data Analysis}
\label{sec:data}

\subsection{REM}
\label{sec:rem}

REM is a 60 cm robotic telescope located at the ESO La Silla observatory (Chile). The telescope simultaneously feeds two cameras (one for the Near-Infrared (NIR) and one for the optical) by a dichroic. The cameras have imaging capabilities with the NIR ($z', J, H$ and $K$) and visual large band ($V, R, I$) filters. 

The main scientific target for REM is the follow-up of the early phases of the infrared and optical afterglow of Gamma-Ray Bursts (GRBs) detected by space-borne high-energy alert systems as Swift\footnote{http://swift.gsfc.nasa.gov}. Apart from GRBs, REM serves the community as a fast pointing imager particularly suited for multi-frequency monitoring of highly variable and transient sources. Among the obvious applications of REM idle time there are AGN and variable stars multi-frequency monitoring. More information about the REM project can be found in \cite{Zerbi01} and \cite{Cov04} and references therein.

\begin{figure}
\includegraphics[width=\columnwidth, height=10.5cm]{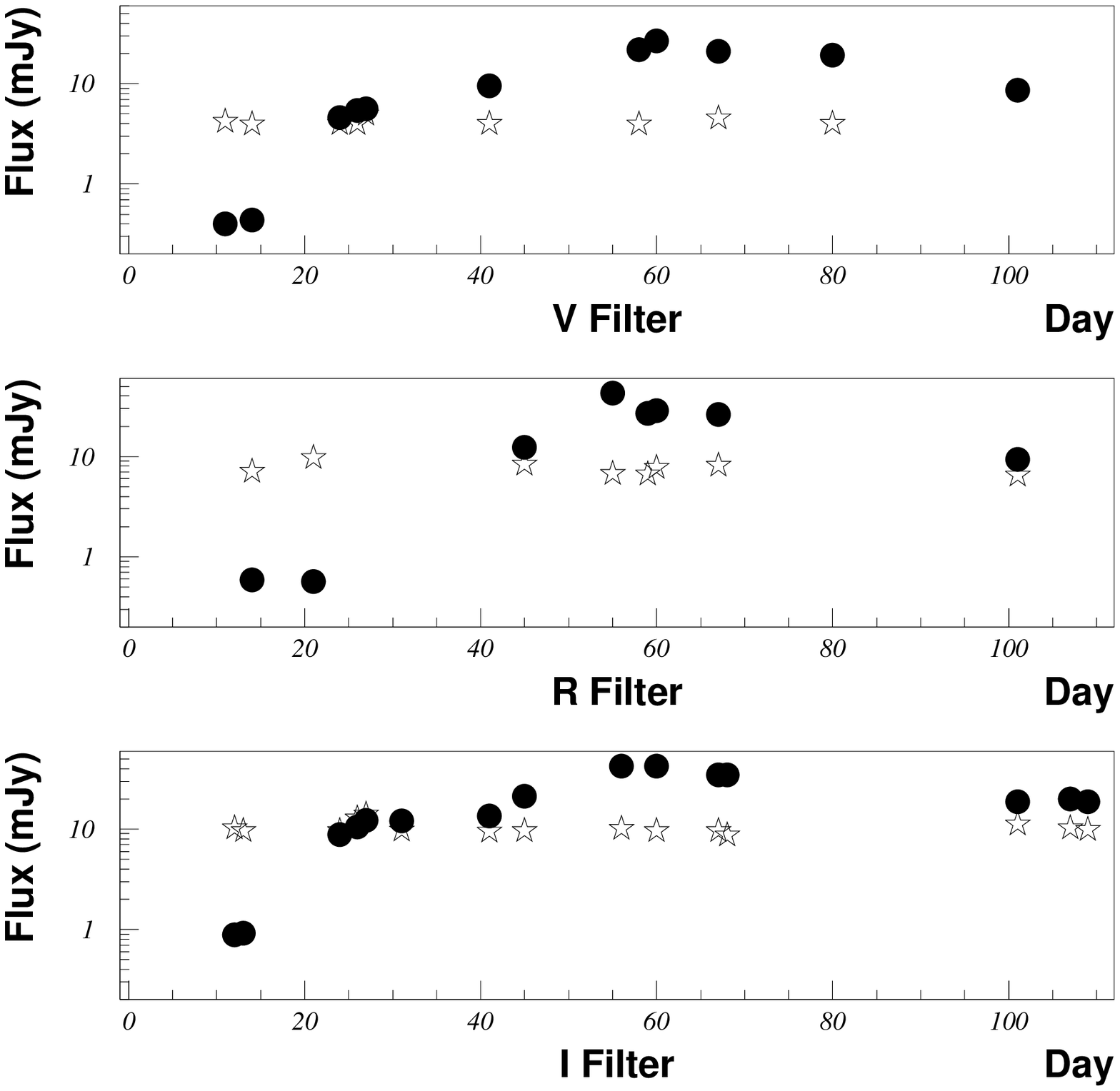} \\
\includegraphics[width=\columnwidth, height=10.5cm]{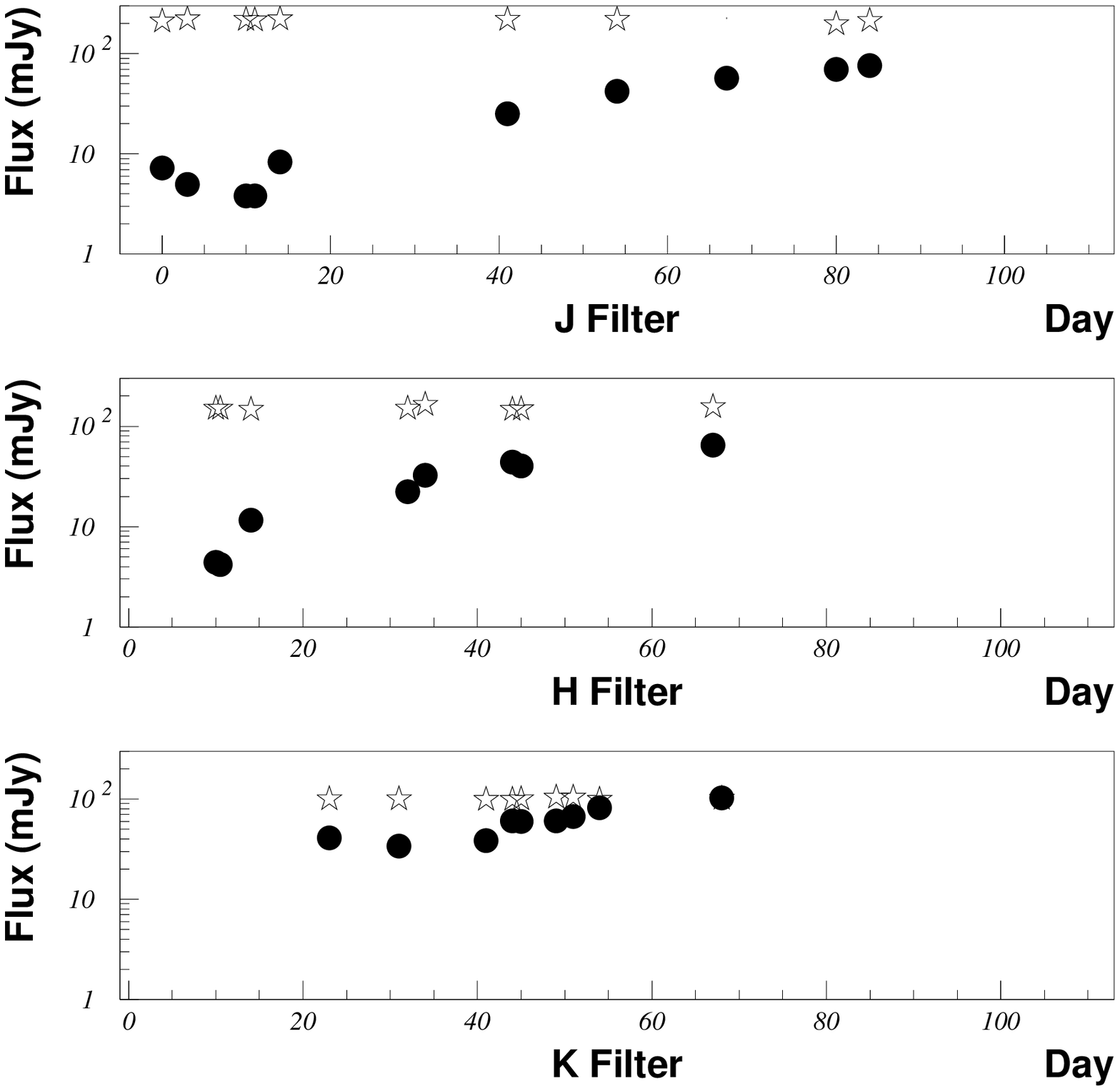}
\caption{Optical and NIR light curves of PKS\,0537$-$441 (filled symbols). The star reported as comparison (open symbols) is one of the six used to perform relative photometry. Photometric errors are comparable to the symbol size. The observation duration for each point is 30 s. Data are reported in Table\,\ref{tab:mag}. Day 0 corresponds to 7 December 2004.}
\label{fig:optirlc}
\end{figure}

\subsection{Observation and Data Analysis}
\label{sec:res}

Photometric observations were obtained from 7 December 2004 to 18 March 2005. Reduction of the REM NIR and optical frames was performed following standard procedures. For photometric analysis we adopted the DAOPHOT package \citep{Stet86} applying standard aperture photometry. Relative calibration for the light curve was obtained using a frame as a reference and computing magnitude shifts to six bright, isolated, not-saturated stars in the field. Absolute calibration was derived in the optical by comparison with photometry obtained for the standard star field PG\,1323$-$086  observed during the same night of the reference frame. Checks on the stability and reliability of the derived zero-points were performed analysing frames obtained during different observing runs. Calibration in the NIR made use of the 2MASS catalogue\footnote{http://irsa.ipac.caltech.edu}. An independent check on the photometric errors was done comparing instrumental magnitudes of non-variable stars in the field obtained in different frames. Relative and absolute calibration errors have been added in quadrature to the photometric errors derived for each object.

The results of the photometric analysis are reported in Table\,\ref{tab:mag} and shown in Fig.\,\ref{fig:optirlc}. Throughout this paper errors are reported at 1$\sigma$ confidence level.

\begin{table}
\begin{tiny}
\centerline{
\begin{tabular}{crcrr}
\hline
\textbf{Date} & \textbf{Exp.} & \textbf{Filter} & \textbf{Mag} & \textbf{Flux} \\
\hline
UT   &  s     &          &        & mJy \\
\hline
07/12/04 05:03:47 & 300 & $J$ & $13.41 \pm 0.05$ & $7.24 \pm 0.41$ \\
10/12/04 07:00:41 & 1500 & $J$ & $13.82 \pm 0.02$ & $4.95 \pm 0.12$ \\
17/12/04 05:35:35 & 1500 & $H$ & $13.18 \pm 0.03$ & $4.37 \pm 0.01$ \\
17/12/04 09:04:28 & 1500 & $H$ & $13.42 \pm 0.02$ & $4.20 \pm 0.19$ \\
17/12/04 04:09:36 & 300 & $J$ & $14.11 \pm 0.04$ & $3.81 \pm 0.17$ \\
18/12/04 04:21:10 & 1500 & $J$ & $14.09 \pm 0.03$ & $3.83 \pm 0.13$ \\
18/12/04 04:12:03 & 600 & $V$ & $17.39 \pm 0.01$ & $0.40 \pm 0.04$ \\
19/12/04 04:34:42 & 600 & $I$ & $16.01 \pm 0.03$ & $0.89 \pm 0.04$ \\
20/12/04 04:14:52 & 600 & $I$ & $15.97 \pm 0.03$ & $0.92 \pm 0.03$ \\
21/12/04 04:17:12 & 1500 & $H$ & $12.32 \pm 0.02$ & $11.53 \pm 0.30$ \\
21/12/04 04:33:17 & 900 & $J$ & $13.26 \pm 0.02$ & $8.27 \pm 0.19$ \\
21/12/04 04:00:16 & 600 & $R$ & $16.72 \pm 0.04$ & $0.59 \pm 0.03$ \\
21/12/04 04:28:11 & 600 & $V$ & $17.31 \pm 0.08$ & $0.44 \pm 0.04$ \\
28/12/04 03:36:59 & 600 & $R$ & $16.76 \pm 0.05$ & $0.57 \pm 0.03$ \\
30/12/04 03:46:53 & 750 & $K$ & $10.45 \pm 0.01$ & $41.08 \pm 0.49$ \\
31/12/04 03:40:03 & 510 & $I$ & $13.52 \pm 0.07$ & $8.78 \pm 0.50$ \\
31/12/04 03:43:57 & 30 & $V$ & $14.75 \pm 0.04$ & $4.56 \pm 0.15$ \\
02/01/05 03:33:04 & 30 & $I$ & $13.33 \pm 0.06$ & $10.48 \pm 0.60$ \\
02/01/05 03:38:36 & 120 & $V$ & $14.57 \pm 0.04$ & $5.39 \pm 0.18$ \\
03/01/05 03:17:43 & 150 & $I$ & $13.16 \pm 0.06$ & $12.23 \pm 0.77$ \\
03/01/05 03:35:51 & 120 & $V$ & $14.52 \pm 0.04$ & $5.65 \pm 0.19$ \\
07/01/05 04:25:33 & 750 & $K$ & $10.66 \pm 0.01$ & $33.82 \pm 0.37$ \\
07/01/05 04:19:30 & 120 & $I$ & $13.17 \pm 0.06$ & $12.13 \pm 0.70$ \\
08/01/05 03:00:49 & 600 & $H$ & $11.60 \pm 0.11$ & $22.41 \pm 2.41$ \\
10/01/05 03:41:32 & 480 & $H$ & $11.20 \pm 0.11$ & $32.45 \pm 3.17$ \\
17/01/05 02:45:02 & 750 & $K$ & $10.52 \pm 0.01$ & $38.44 \pm 0.43$ \\
17/01/05 02:53:29 & 750 & $J$ & $12.06 \pm 0.01$ & $25.02 \pm 0.30$ \\
17/01/05 02:31:47 & 150 & $I$ & $13.05 \pm 0.06$ & $13.55 \pm 0.78$ \\
17/01/05 02:47:23 & 150 & $V$ & $13.96 \pm 0.04$ & $9.53 \pm 0.32$ \\
20/01/05 02:18:51 & 630 & $K$ & $10.03 \pm 0.02$ & $60.09 \pm 1.00$ \\
20/01/05 02:30:14 & 300 & $H$ & $10.87 \pm 0.11$ & $44.18 \pm 4.27$ \\
21/01/05 02:21:13 & 570 & $K$ & $10.15 \pm 0.01$ & $59.95 \pm 0.60$ \\
21/01/05 02:29:42 & 750 & $H$ & $10.97 \pm 0.11$ & $40.29 \pm 3.90$ \\
21/01/05 02:09:33 & 120 & $I$ & $12.56 \pm 0.07$ & $21.29 \pm 1.24$ \\
21/01/05 02:24:10 & 120 & $R$ & $13.42 \pm 0.07$ & $12.32 \pm 0.80$ \\
25/01/05 02:05:43 & 570 & $K$ & $10.03 \pm 0.02$ & $60.09 \pm 1.22$ \\
27/01/05 01:53:15 & 1470 & $K$ & $9.92 \pm 0.02$ & $66.62 \pm 0.92$ \\
30/01/05 01:38:50 & 420 & $K$ & $9.67 \pm 0.01$ & $82.03 \pm 0.98$ \\
30/01/05 01:44:15 & 750 & $J$ & $11.50 \pm 0.01$ & $41.87 \pm 0.50$ \\
31/01/05 01:31:25 & 270 & $R$ & $12.06 \pm 0.07$ & $43.00 \pm 2.85$ \\
01/02/05 01:20:59 & 120 & $I$ & $11.82 \pm 0.06$ & $42.29 \pm 2.42$ \\
03/02/05 01:21:42 & 120 & $V$ & $13.05 \pm 0.04$ & $21.89 \pm 0.73$ \\
04/02/05 01:08:30 & 120 & $R$ & $12.56 \pm 0.07$ & $27.06 \pm 1.76$ \\
05/02/05 02:43:21 & 150 & $I$ & $11.81 \pm 0.06$ & $42.52 \pm 2.56$ \\
05/02/05 02:47:51 & 150 & $R$ & $12.50 \pm 0.07$ & $28.81 \pm 1.88$ \\
05/02/05 02:51:41 & 120 & $V$ & $12.83 \pm 0.04$ & $26.76 \pm 0.89$ \\
12/02/05 01:18:11 & 720 & $H$ & $10.45 \pm 0.11$ & $64.87 \pm 6.27$ \\
12/02/05 01:33:29 & 750 & $J$ & $11.17 \pm 0.02$ & $56.69 \pm 0.78$ \\
12/02/05 00:50:37 & 150 & $I$ & $12.03 \pm 0.07$ & $34.82 \pm 2.00$ \\
12/02/05 01:12:10 & 150 & $R$ & $12.60 \pm 0.07$ & $26.22 \pm 1.71$ \\
12/02/05 01:27:49 & 120 & $V$ & $13.10 \pm 0.04$ & $20.95 \pm 0.86$ \\
13/02/05 01:40:39 & 540 & $K$ & $9.45 \pm 0.01$ & $102.99 \pm 1.14$ \\
13/02/05 01:34:43 & 120 & $I$ & $12.03 \pm 0.06$ & $34.82 \pm 2.00$ \\
18/03/05 00:41:24 & 150 & $I$ & $12.69 \pm 0.11$ & $18.84 \pm 1.93$ \\
18/03/05 00:45:49 & 150 & $R$ & $13.72 \pm 0.07$ & $9.36 \pm 0.61$ \\
24/03/05 00:41:15 & 150 & $I$ & $12.62 \pm 0.06$ & $20.07 \pm 1.16$ \\
25/02/05 02:12:19 & 750 & $J$ & $10.96 \pm 0.02$ & $69.23 \pm 0.96$ \\
25/02/05 02:04:58 & 120 & $V$ & $13.19 \pm 0.04$ & $19.26 \pm 0.65$ \\
26/03/05 00:42:30 & 120 & $I$ & $12.70 \pm 0.06$ & $18.72 \pm 1.07$ \\
01/03/05 01:38:10 & 750 & $J$ & $10.85 \pm 0.01$ & $76.40 \pm 0.98$ \\
18/03/05 00:49:20 & 120 & $V$ & $14.06 \pm 0.04$ & $8.68 \pm 0.29$ \\
\hline
\end{tabular} 
}
\end{tiny}
\caption{Photometry of PKS\,0537$-$441 from December 2004 to March 2005. The UT corresponds to the middle of the exposures. The gross exposure time is reported.}
\label{tab:mag}
\end{table}

\begin{figure}
\includegraphics[width=\columnwidth, height=15cm]{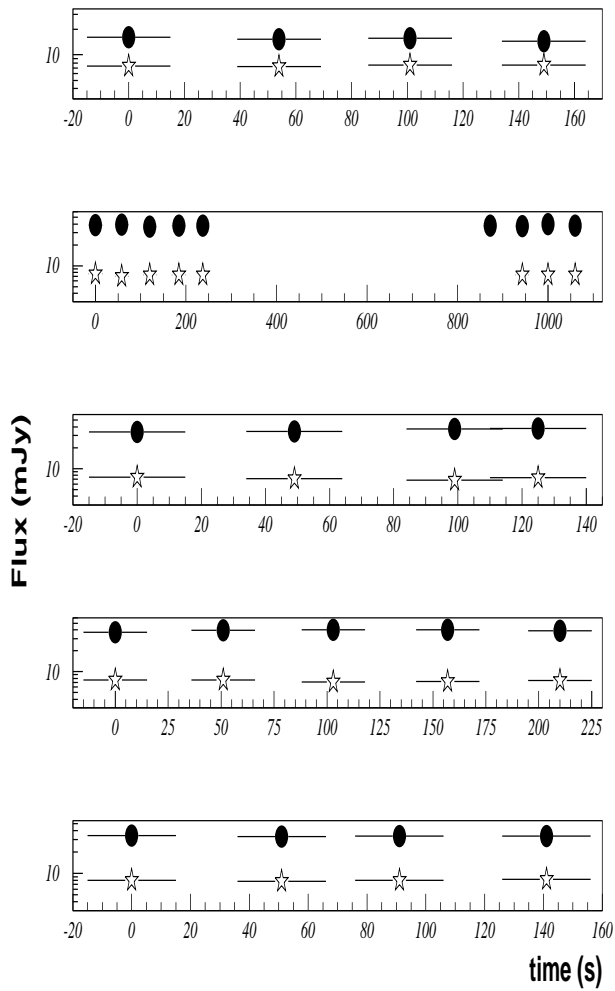} \\
\caption{$R$ filter short-term variability at five epochs for PKS\,0537$-$441 (filled symbols).  From top to bottom: 21 January, 31 January, 4 February, 5 February and 12 February 2005. Time 0 corresponds to the beginning of the observing sequence. The star reported as comparison (open symbols) is one of the six used to perform relative photometry. No clear evidence of short-term variability is detected although a weak sign of higher variability for PKS\,0537$-$441 compared to the reference star is present. Photometric errors are comparable to the symbol size, observation duration, 30 seconds, is reported as an horizontal bar superposed on the symbols.}
\label{fig:micro}
\end{figure}

\begin{figure}
\includegraphics[width=6.5cm, angle=-90]{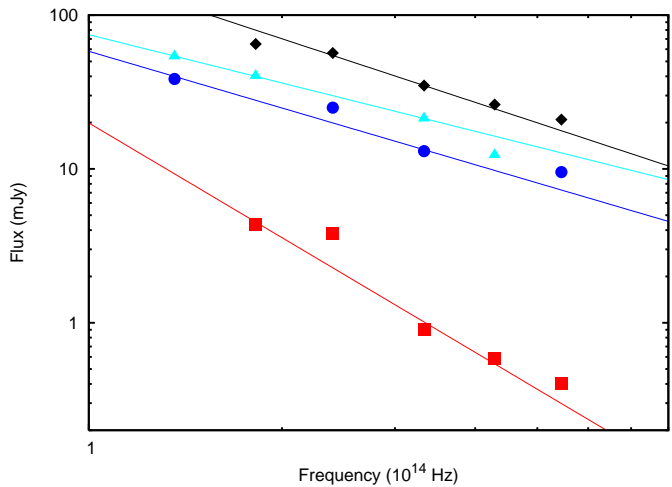}
\caption{Spectral flux distributions at four epochs. For the first epoch, in December 2004, the observations spanned from 17 December to 21 (squares). On 17 January 2005 (circles), 21 January 2005 (triangles), and 21 February 2005 (diamonds) the observation were carried out within one hour. For the errors see Table\,\ref{tab:mag}. The spectral slopes, $F(\nu) \propto \nu^{-\alpha}$, are $\alpha = 2.48 \pm 0.32$, $1.22 \pm 0.17$, $1.04 \pm 0.04$, $1.37 \pm 0.12$, respectively.}
\label{fig:seds}
\end{figure}

\section{Results and Discussion}
\label{sec:disc}

Our photometry clearly shows a variation of almost a factor 100 in the V band in 50 days, and of about a factor 10 in 10 days. The maximum brightness occurred at the end of January 2005. Some decrease in the optical is apparent thereafter, while in the J band the flux continues to rise.  The more limited sampling in H and K bands prevents us to confirm this different trend observed in optical and infrared (Fig.\,\ref{fig:optirlc}).

As mentioned in Sect.\,\ref{sec:pks} active phases are not rare for the source. \citet{Egg73} reported an episode of about four magnitudes in V from the September 1971 to May 1972. Although the amplitude of our event and its duration is similar to that of \citet{Egg73}, we note that our coverage contains much more data and in six photometric bands. 

No clear evidence of variability within a single night was found even if some hints may be present (see Fig.\,\ref{fig:micro} and in particular the first box). In fact the percent variation is more than twice that of the reference star (2.2\% vs. 0.9\%). This indicates the interest of performing photometry of several hours duration, intermediate between one night and a few minutes.

We fitted our photometric points (weighted by the uncertainties) with a power-law at various levels of intensity. The fits are not always very satisfactory possibly because the photometries obtained in the various bands were not strictly simultaneous (Fig.\,\ref{fig:seds}). The low-state of December 2004 clearly corresponds to the softest state, a behaviour which is common in Blazars and specifically in low-frequency peaked Blazars \citep{Pad95,Foss98}. On the other hand, the observations of January 17 and February 21 exhibit within the errors the same slope, while the intensity varied by about 0.5 mag.

This combination of chromatic and achromatic variability behaviours is not unusual in Blazars. The optical monitoring of Blazars at time scales of hours to minutes has detected remarkable flares \citep{Mat99, Dai01, Vil02, ReP03,  Fuh05, Wu05}.  The simultaneous spectral variations are some times clearly correlated with the flux changes, but most often there is no clear correlation, so that one possible conclusion is that both intrinsic mechanisms \citep[related to the propagation of a shock in the jet,][]{MeT96, Spa01} and geometrical effects \citep[variation of relativistic beaming factor,][]{VeR99} may concur to produce the observed variability and the complex relationship between spectrum and flux variations.  Very accurate, intensive, and regular monitorings are needed to disentangle the different components and estimate the significance of each contribution.  In addition, clarifying the state-dependence of the variability character may lead to an understanding of the occasionally complicated correlation between variations at optical and high frequencies \citep[see][]{ReP03}.

The photometric system of REM is now fully operational, and photometric output can be provided in real-time. Were this the case during  our observations of December 2004 - March 2005, our photometry could have triggered observations in the X- and gamma-bands. In that case the evolution of the spectral flux distribution could have been followed over a broad band, which is the best test for emission model available so far, and in particular for clarifying the meaning of achromatric and chromatric variability in the optical band. 
Because of its brightness and luminosity ($z \sim 1$) PKS\,0537$-$441 remains an ideal candidate for this kind of studies.

\begin{acknowledgements}
REM is a collaborative project of Italian Observatories and Universities funded by the Ministry for Education, Universities and Research. We also gratefully acknowledge a special grant by INAF (Rome) to operate the REM telescope at ESO (La Silla). AFS acknowledges support from the Spanish Ministry of Education project AYA2002-03326 and the Generalitat Valenciana project GV/GRUPOS03/170. We also thank the anonymous referee for comments and suggestions which helped us to improve the paper.
\end{acknowledgements}

\end{document}